\documentclass[a4paper]{report}
\usepackage[utf8]{inputenc}
\usepackage[T1]{fontenc}
\usepackage{RJournal}
\usepackage{amsmath,amssymb,array}
\usepackage{booktabs}

\newlength{\cslhangindent}
\newlength{\csllabelwidth}
\newlength{\cslentryspacingunit} 
  {}%
  {\par}
 {
  \setlength{\parindent}{0pt}
  \ifodd #1
  \let\oldpar\par
  \def\par{\hangindent=\cslhangindent\oldpar}
  \fi
  \setlength{\parskip}{#2\cslentryspacingunit}
 }%
 {}
\usepackage{calc}

\usepackage{algorithm}
\usepackage{algorithmic}

\begin{document}

\sectionhead{Contributed research article}
\volume{XX}
\volnumber{YY}
\year{20ZZ}
\month{AAAA}

\begin{article}
\title{ORKM: An R Package for Online Multi-View Data}
\author{by Miao Yu and Guangbao Guo}

\maketitle

\abstract{%
\textcolor{blue}{We introduce a software package, denoted as \pkg {ORKM}, that incorporates the Online Regularized K-Means Clustering (ORKMC) algorithm for processing online multi/single-view data. The function \code{ORKMeans} of the ORKMC utilizes a regularization term to address multi-view clustering problems with online updates. The package \pkg{ORKM}  is capable of computing the classification results, cluster center matrices, and  weights for each view of the multi-view data sets. Furthermore, it can handle branch multi/single-view data by transforming the online RKMC algorithm into an offline version, referred to as Regularized K-Means Clustering (RKMC). We demonstrate the effectiveness of the package through simulations and real data analysis, comparing it with several methods and related R packages. Our results indicate that the package is stable and produces good clustering outcomes.}
}

\hypertarget{introduction}{%
\section{Introduction}\label{introduction}}

\hypertarget{motivation}{%
\subsection{Motivation}\label{motivation}}

K-means \citep{MacQueen1967,MKM,softKM} as one of the popular clustering methods, has a wide range of applications in various fields. Its objective function of single-view K-means can be expressed as \[\min _{C_{K}}\sum_{k=1}^{K} \sum_{X_{i\cdot} \in C_{k}} \|X_{i\cdot}-M_{k \cdot}\|^2,\] where \(X=(X_{i j}) \in \mathbb{R}^{N \times J}\) is an input data matrix, \(X_{i\cdot}\) is the \(i\)-th row-vector of data matrix \(X\), \(M=(M_{k j}) \in \mathbb{R}^{K \times J}\) is a center matrix with the \(k\)-th cluster center vector \(M_{k \cdot}\in \mathbb{R}^{1 \times J}\), \(C_{k}\) is the \(k\)-th cluster, \(K\) is the number of cluster, \(N\) is the size of samples, and \(J\) is the number of features.

\textcolor{blue}{There exist numerous multi-view data sets in various fields, such as information processing, natural language, internet, biology, among others. In these fields, each feature of the data can be considered a view of the data set. It is important to note that in certain cases, single view data may not adequately represent the characteristics of cluster data sets, and thus the use of multi-view data is becoming increasingly prevalent.} For a data set including \(V\) views, the multi-view data matrix \(X=(X_{ij})=(X^{1}, X^{2}, \ldots, X^{V}) \in \mathbb{R}^{N \times J_{V}}\) is an input data matrix on the view data matrix sequence \(\{X^{v}=(X_{ij}^v)_{N\times J}\in \mathbb{R}^{N\times J}\}_{v=1}^V\) where \(N\) is the number of samples, \(J\) is the column number of the data matrix \(X\). Without all these features of these views, the accuracy of data representation will be questioned. By using non-negative matrix factorization, we have \[
X^v\approx UM^v,
\] where \(X^{v}=(X_{i j}^{v}) \in \mathbb{R}^{N \times J}\) is an input data matrix on the \(v\)-th view, \(M^{v}=(M_{k j}^{v}) \in \mathbb{R}^{K \times J}\) is a center matrix on the \(v\)-th view, and \(U=(U_{ik}) \in \mathbb{R}^{N \times K}\) is a clustering indicator matrix, \(X_{i\cdot}^{v}\) is the \(i\)-th row of the \(v\)-th data matrix \(X^{v}\). If \(X_{i \cdot}^{v}\) belongs to \(C_{k}\), then \(U_{ik}=1\); otherwise, \(U_{ik}=0\). To obtain a good approximation of the matrix \(X^v\) , it is necessary to evaluate the reconstruction error between the matrix \(X^v\) and the matrix \(UM^v\), so we use F-norm to calculate the error, which is expressed as, \[\|X^{v}-UM^{v }\|^{2}_{F}.\]

We use nonnegative matrix factorization(NMF) to express the objective function of K-means as \[ \sum_{v=1}^{V} \|X^{v}-UM^{v  }\|^2_{F} \ {\rm for} \sum_{k=1}^{K} U_{ik}=1, U \geq 0, M^{v} \geq 0.\]

We often add regular terms to the objective function, called regularization, and reduce the possibility of overfitting. The objective function of regularized K-means clustering(RKMC) for multi-view data, can be expressed as \[J_{\rm RK}=  \sum_{v=1}^{V} \|X^{v}-UM^{v }\|^2_{F} + \eta \operatorname{trace}(U U^{\top}) \ {\rm for} \sum_{k=1}^{K} U_{ik}=1, U \geq 0, M^{v} \geq 0,\] where \(\eta\) is the regularization parameter. We can solve the optimization problem through Lagrange multiplier method. The Lagrange function \(\mathcal{L}\left(U_{ik},\lambda\right)\) is defined as \[\mathcal{L}(U_{ik}, \lambda)=J_{\rm RK}-\lambda(\sum_{i=1}^{N} \sum_{k=1}^{K} U_{i k}-1),
\] where \(\lambda\) is the Lagrange multiplier. Taking the derivative of the function \(\mathcal{L}(U_{i k},\lambda)\) with respect to \(U_{ik}\) to zero.

\hypertarget{online-multi-view-data-problem}{%
\subsection{Online multi-view data problem}\label{online-multi-view-data-problem}}

\textcolor{blue}{A clustering method that exhibits high performance on training sets but produces overlapping results for real data sets will not be applicable in dealing with the latest data sets. Recently, online data has become increasingly prominent due to the global impact of COVID-19. Online education and healthcare have become common practices, while stock market data are updated in real time. Additionally, research efforts have shifted from single view data to multi view data. Consequently, developing effective strategies for handling online multi view datasets poses a significant theoretical challenge.}

\textcolor{blue}{The exploration of the data dimension in current data sets has led to an increase in computational costs for the presented algorithms, as well as high memory costs for data storage. Offline clustering methods cannot process online data sets, such as stock data, financial transaction data, and online education data. However, the online method for multi-view data is capable of fulfilling these requirements.} In online multi-view data set, the data matrix \(X^{v}=(X_{ij}^{v})_{N \times J}=(X_{1}^{v \top}, \ldots, X_{t-1}^{v \top} ,X_{t}^{v \top}, \ldots, X_N^{v \top})^\top\in \mathbb{R}^{N \times J}\) is the data matrix of the \(v\)-th view. In online form, at time \(t\), the online multi-view data matrix \(X^{v (t)}=(X_{ij}^{v})_{t \times J} =(X^{v(t-1)\top} ,X_{t}^{v \top} )^\top \in \mathbb{R}^{t\times J}\) divided into two parts, where the matrix \(X^{v (t-1)} =(X_1^{v \top},\ldots,X_{t-1}^{v \top})^\top =(X_{ij}^{v})_{(t-1) \times J}\in \mathbb{R}^{(t-1) \times J}\) is all data received before time \(t\), and the vector \(X_{t}^{v}=(X_{tj}^{v})\in \mathbb{R}^{1 \times J}\) is data received at time \(t\). So the objective function of online multi-view data can be expressed as \[
\sum_{v=1}^{V} \sum_{i=1}^{t} (\alpha^{v (t)})^{r} \|X^{v(t)}-U^{(t)} M^{v  (t) }\|^2_{F} \ {\rm for} \sum_{k=1}^{K} U_{ik}^{ (t)}=1, U^{(t)} \geq 0, M^{v  (t)} \geq 0,
\] where the matrix \(M^{v(t)}=(M_{kj}^{v(t)}) \in \mathbb{R}^{K \times J}\) is the \(k\)-th center matrix at time \(t\). The matrix \(U^{(t)}=(U_{ik }^{(t)}) \in \mathbb{R}^{t \times K}\) is the indicator matrix for \(X^{v (t)}\) and \(M^{v (t)}\) with \(U_{ik }^{(t)} \in \left\{0,1\right\}\). \(\alpha^{v(t)}\) and \(r\) are the weight parameter of \(X^{v(t)}\) in multi-view data .

Let \(v=1\) in online multi-view data matrix \(X^{v (t)}\), we can get a single-view data matrix \(X^{(t)}\). Therefore, we can summarize the basic elements of online multi-view data problems as follows. The cluster \(C_{k}(k=1,\ldots,K)\) for the online data matrix \(X^{v (t)}\); the indicator matrix \(U^{(t)}\) for \(X^{v (t)}\); the cluster center matrix \(M^{v (t)}\) of cluster \(C_{k}\) on \(U^{(t)}\); the weight of each view \(\alpha ^{v(t)}\) on \(U^{(t)}\) and \(M^{v (t)}\). Note that the computation accurate of \(U^{(t)}\) can make the clustering effect better, and the appropriate regularization term can prevent from overfitting, the most important problems among them are that how to compute \(U^{(t)}\) for \(X^{v (t)}\), and choose the regularization term.

\hypertarget{related-developments}{%
\subsection{Related developments}\label{related-developments}}

Clustering is one of the most important methods in multi-view data sets. Note that K-means was firstly proposed by MacQueen \citep{MacQueen1967}, Ding \citep{Ding2005} proposed orthogonal non-negative matrix factorization (NMF) for K-means clustering.

\textcolor{blue}{Existing multi-view clustering methods fall into two categories: generative (or model-based) methods and discriminative (or similarity-based) methods. based) methods. Generative methods attempt to learn the underlying distribution of the data and represent the data using generative models, each representing a cluster. Each model represents a cluster. Discriminative methods further divide them into five categories based on how they combine multi-view information \citep{chao2018}: common feature vector matrix (multi-view atlas clustering \citep{kumar2011co}, \citep{lu2016}), common coefficient matrix (multi-view subspace clustering \citep{white2012}, \citep{zhao2014}), common indicator matrix (multi-view non-negative matrix (multi-view non-negative matrix decomposition clustering \citep{zong2017}, \citep{DMC}, \citep{OMU}), direct view combination (multi-core clustering \citep{kumar2011co}), and post-projection view combination (typical correlation analysis). The first three classes have one thing in common, they share a similar structure to combine multiple views. Cai \citep{MKM} proposed multi-view K-means clustering via non-negative matrix factorization.}

\textcolor{blue}{For online clustering, Liberty \citep{OKM} presented an online K-means; Yang \citep{yang2021a} presented an online binary incomplete multi-view clustering; Shao \citep{shao2016b} proposed an online multi-view clustering, which deals with large-scale incomplete views. Guo et al. proposed an incremental limit learner for online sequential learning problems \citep{guo2014incremental}.  Kim et al. proposed an optimisation method for online update regularisation \citep{kim2018}. Zhou et al. proposed an online regular kernel limit learning method with forgetting mechanism \citep{zhou2014}. Lee et al. proposed an online correction method for measurement errors in large data streams \citep{lee2020}.} For regularization terms, we can select them with Lasso, adaptive Lasso \citep{zou2006adaptive}, Elastic net \citep{zou2005regularization}, SCAD \citep{wang2007group}, MCP \citep{zhang2010nearly}, and so on. In particular, co-regularization \citep{kumar2011co} can also learn information from different views of multi-view data sets.

There are many researchers who have uploaded the packages of K-means clustering in the R website, most of which are suitable to deal with single-view data sets. For example, \CRANpkg{ClusterR} \citep{ClusterR} includes K-Means, Mini Batch K-means, and so on while \CRANpkg{mclust} \citep{mclust} presents model-based clustering in Gaussian finite mixture model. \CRANpkg{Spectrum} \citep{Spectrum} presents a fast adaptive spectral clustering in multi/single-view data sets.

\hypertarget{our-work}{%
\subsection{Our work}\label{our-work}}

\textcolor{blue}{Despite the numerous advantages of the current clustering algorithm for online multi-view data, it has some limitations. Firstly, it is unable to process online data. Secondly, the performance of the online clustering algorithm falls short of expectation.} Here we propose the package \pkg{ORKM} \citep{ORKM} with the functions including \code{ORKMeans}, \code{RKMeans}, and \code{INDEX}. The function \code{RKMeans} is another major function without online update. \textcolor{blue}{The existing clustering algorithms for online multi-view data have several advantages; however, they fail to process online data or their performance is insufficient. To address this issue, we propose a novel clustering algorithm based on normalized multi-information (NMI), purity, and F-score as performance indicators. Our approach overcomes these limitations by effectively processing online multi-view data while also achieving better clustering performance than other algorithms and similar R packages. The highlights of our package are as follows.}

\begin{itemize}
  \item The function \code{ORKMeans} can deal with online multi/single-view data sets, we can set the initial value of the online method via  \code{chushi=}.
      \item Both  \code{ORKMeans} and \code{RKMeans} can deal with   multi/single-view data sets, we can use \code{V=1} or \code{V=} to control.
    \item \code{ORKMeans} and \code{RKMeans} give the clustering results by \code{\$result}. We also give the weight value in each view of  multi-view data sets by \code{\$weight}.
   \item The functions \code{ORKMeans} and \code{RKMeans}  have a regularization term to reduce over-fitting. Note that the regularization parameters are different in different data sets, we can select the regularization parameters through \code{yita=}.
       \item We have written the index values for the clustering effect as a function \code{INDEX}.  We  have also given  three indicators (RI, Precision, and Recall) to evaluate the clustering effect.
  \item The other five methods we have compared to the ORKMC have also been written as the functions, including  \code{KMeans} \citep{MKM,softKM}, \code{PKMeans} \citep{PKM}, \code{OGD} \citep{OGD}, \code{OMU} \citep{OMU}, and \code{DMC} \citep{DMC}.
\end{itemize}

The rest of this paper is structured as follows. Section 2 presents the proposed method to deal with online multi-view data problems while Section 3 presents the R package and the comparison package. Section 4 presents how to use package 5 while Section 6 presents the results of its sensitivity and stability. Section 7 summarizes the advantages and disadvantages of package \pkg{ORKM}.

\hypertarget{method}{%
\section{Method}\label{method}}

Note that multi-view data often arrives in the form of online update, how to handle the clustering problem of such online multi-view data is another important challenge. The objective function of online regularized K-means clustering(ORKMC) for online multi-view data can be expressed as

\[
J_{ON}= \sum_{v=1}^{V} \sum_{i=1}^{t} (\alpha^{v (t)})^{r} \|X^{v (t)}-U^{(t)}M^{v (t) }\|^2_{F} + \eta \operatorname{trace}(U^{(t)} U^{ (t)\top})) \ {\rm for} \sum_{k=1}^{K} U_{ik }^{(t)}=1, U^{(t)} \geq 0, M^{v(t)} \geq 0,
\] where \(\eta\) is the regularization parameter, \(X^{v(t)}\) is the input data chunk of the \(v\)-th view before time \(t\), and \(M^{v(t)}\) is the center matrix of the \(v\)-th view before time \(t\). \(\alpha^{v(t)}\) is the weight parameter of \(X^{v (t)}\) in multi-view data before time \(t\), \(r\) is a constant.

We approximate the optimal solution by processing the data one by one or chunk by chunk. Note that the objective function is not convex with \(U^{(t)}\) and \(M^{v(t)}\), the above minimization problem is difficult to be solved. Many common solutions are presented, by multiplicative updates \citep{lee1999learning}, project gradient descent, stochastic gradient descent \citep{feng2013online}, and incremental method \citep{guo2014incremental}. For the step length \(\gamma\) and the project matrix \(P\), with the project gradient decent, the update equation for \(U^{(t)}\) is expressed as

\begin{equation} \label{onlineu}
\begin{aligned}
U^{(t+1)}&= P\left[U^{(t)}-\gamma \frac{\partial {J_{ON}}}{\partial {U^{(t)}}}\right]. \\
\end{aligned}
\end{equation}

The ORKMC for multi-view data is summarized in Algorithm \ref{Alg2}.

\begin{algorithm}[htbp]
\renewcommand{\algorithmicrequire}{\textbf{Input:}}
\renewcommand{\algorithmicensure}{\textbf{Output:}}
  \caption{The online regularized $K$-means clustering for online multi-view data}\label{Alg2}
  \begin{algorithmic}[1]
    \REQUIRE \emph{Data matrix  $X^{v(t)}$ , clustering $K$, parameters $\eta$, $r$, step length $\gamma$.}
    \STATE \emph{Initialized: random center matrix $M^{v(t)}$, the random cluster indicator matrix $U$, the weight factor $\alpha^{v(t)}=1/V$  for each view.}
    \FOR {$i=t+1,\ldots, N$}
    \STATE \emph{Update the cluster indicator vector $U^{(i+1)} $ for each view }
    \STATE \emph{Update the centroid matrix $M^{v(i+1)}$ for each view  }
    \STATE \emph{Update  $\alpha^{v(i+1)}$ for each view }
    \ENDFOR
    \ENSURE  \emph{Cluster indicator matrix $U \in \mathbb{R}^{N\times K}$, center matrix $M^{v } \in \mathbb{R}^{K\times J}$.}
  \end{algorithmic}
\end{algorithm}

\textcolor{blue}{In this study, we introduce the ORKMC algorithm as a solution to the online multi-view data problem. The proposed approach employs K-means clustering and incorporates regularization terms to mitigate overfitting. Specifically, we utilize RKMeans function in the package to implement the regularized K-means clustering (RKMC) approach. Additionally, we employ projected gradient descent method to resolve the online multi-view data challenge, resulting in the formulation of the ORKMC algorithm that leverages ORKMeans function. This novel algorithmic framework is aimed at addressing challenges related to handling multi-view data, while simultaneously avoiding overfitting, thereby enhancing its accuracy and generalizability.}

\hypertarget{comparison}{%
\section{Comparison}\label{comparison}}

\textcolor{blue}{We present four functions of comparison methods, which are shown in Table \ref{table1a}. And we  presents some introductions to the functions and the R packages.}

\begin{table}[htp] \small
     \begin{center}
     \caption{The comparison functions in the package \CRANpkg{ORKM} }\label{table1a}
\resizebox{\textwidth}{!}{
  \begin{tabular}{ c c c c c c}
  \hline
     Comparison function & Detail & Single-view & Multi-view & Online & Reference\\\hline
     \code{PKMeans} & power K-means clustering & $\surd$ & & & \citep{PKM} \\
     \code{OGD} & online gradient descent & $\surd$ & & $\surd$ & \citep{OGD} \\
     \code{OMU} & online multiplicative update &  & $\surd$ & $\surd$ & \citep{OMU} \\
     \code{DMC} & deep matrix clustering &  &  $\surd$ & & \citep{DMC} \\
     \hline
   \end{tabular}}
   \end{center}
\end{table}

\textcolor{blue}{\textbf{PKMeans:}
The power of K-means smooth intermediate problems are associated with objective functions that are themselves smoother and consequently facilitate the identification of a global minimum consistent with the K-means objective. The algorithm retains the same iteration complexity as Lloyd's algorithm, yet it increases robustness to initialization and significantly enhances performance in high-dimensional clustering scenarios.}

\textcolor{blue}{\textbf{DMC:}
The proposed DMC algorithm performs a layer-by-layer decomposition of the multi-view data matrix into representative subspaces and generates a cluster at each level. To simultaneously optimize for quality and diversity, an iterative optimization process is introduced to seek multiple clusters.}

\textcolor{blue}{\textbf{KMeans:}
The K-means algorithm is an unsupervised learning method in which categories are not predetermined, and similar objects are grouped together automatically. Clustering is performed by computing the distances between each data point and the center of mass of various clusters, followed by assignment to the closest cluster. Its simplicity and ease of implementation are among its strengths.}

\textcolor{blue}{\textbf{OMU:}
The proposed method involves incorporating a multiplicative normalization factor as an additional term into the original additivity update rule. Typically, this additional term has an approximately opposite direction to the original rule. At the end of each iteration, non-negativity is ensured and maintained.}

\textcolor{blue}{\textbf{OGD:}
The present algorithm employs the gradient of the current sample to determine the direction of each update. The locality or globality of the extremes in the gradient descent method relies on the sequence of sampled instances. }

To verify the clustering performance of the package \CRANpkg{ORKM}, we also present other cluster-related packages as comparison packages, see also Table \ref{table1b}.

\begin{table}[htp] \small
 \begin{center}
\caption{The comparison packages}\label{table1b}
\resizebox{\textwidth}{!}{
  \begin{tabular}{ c c c c c}
       \hline
     Comparison packages & Detail & Single-view & Multi-view & Reference \\ \hline
     \CRANpkg{ClusterR} &  Mini-Batch-Kmeans & $\surd$ & & \citep{ClusterR} \\
     \CRANpkg{mclust}   &  model-based clustering  & $\surd$ &   & \citep{mclust} \\
     \CRANpkg{Spectrum} & fast adaptive spectral clustering & $\surd$ &  $\surd$ & \citep{Spectrum} \\
     stats & R statistical functions &$\surd$ & & (R-core, 2023) \\
     \CRANpkg{fpc} & Various methods for clustering and cluster validation & $\surd$& &\citep{fpc} \\
     \hline
   \end{tabular}}
   \end{center}
\end{table}

\textcolor{blue}{\textbf{ClusterR:}
The \CRANpkg{ClusterR} package facilitates the application of K-means and block K-means clustering within a Gaussian mixture model. It features various functionalities, such as cluster visualization, validation, prediction for new data, and estimation of the ideal number of clusters.}

\textcolor{blue}{\textbf{mclust:}
The \CRANpkg{mclust} utilizes the EM algorithm to fit a Gaussian mixture model. It offers precise control over both size and shape of the covariance matrix while also offering hierarchical clustering through employment of maximum likelihood methods. Moreover, it employs an integrated clustering approach using Bayesian Information Criterion (BIC) and density estimation, in addition to discriminant analysis, to achieve optimal results.}

\textcolor{blue}{\textbf{Spectrum:}
The \CRANpkg{Spectrum} package provides an adaptive method for spectral clustering of multi-view data, which incorporates a multimodal heuristic that effectively determines the number of clusters. The proposed approach offers significant flexibility by automatically selecting the cluster number and accommodating diverse Gaussian and non-Gaussian structures during the clustering process.}

\textcolor{blue}{\textbf{stats:}
The stats is the most important statistical package in R. It contains many basic functions such as cluster analysis, time series analysis, model fitting and many more.}

\textcolor{blue}{\textbf{fpc:}
The \CRANpkg{fpc} is one of the most specialized packages in R for cluster analysis, and it contains various cluster analysis and cluster validation methods. For example, K-means, etc.}

\hypertarget{usage}{%
\section{Usage}\label{usage}}

In this subsection, we mainly describe the implementation of the functions \code{ORKMeans} and \code{RKMeans}. We install and load the latest version of the package from CRAN as follows,

\begin{verbatim}
install.packages("ORKM")
library(ORKM)
\end{verbatim}

With several given parameters, the package can complete the work of cluster analysis. For example, we set the initial value ``\code{chushi}'', the number of clusters ``\code{K}'', the balance parameter ``\code{r}'', the regularization parameter ``\code{yita}'', the step size ``\code{gamma}'', the stop threshold ``\code{epsilon}'', the true label ``\code{truere}'', the maximum number of iterations ``\code{max.iter}'', and the NMI ``\code{method=0}'', expressed as

\begin{verbatim}
set.seed(1234)
ORKMeans (X = X, K = K, V = V, chushi = chushi, r = r, yita = yita,
   + gamma = gamma, epsilon = epsilon, max.iter = max.iter, truere = truere, method = 0)
RKMeans (X = X, K = K, V = V, yita = yita, r = r, max.iter = max.iter,
   + truere = truelabel, method = 0)
\end{verbatim}

With \code{ORKMeans} and \code{RKMeans}, we have some output values, shown in Table \ref{tabou}. Note that the parameters are different in different data sets, we need to enter the appropriate parameters.

\begin{table}[htp] \small
 \begin{center}
  \caption{The output indicators of \code{ORKMeans} and \code{RKMeans} in the package \CRANpkg{ORKM} } \label{tabou}
  \begin{tabular}{ c c c }  \hline
 \code{ORKMeans} & \code{RKMeans} & Description \\ \hline
  NMI & NMI &   The result of  NMI indicator  \\
  weight & weight  &  Weight  results of different views\\
   center& center & The center matrix result \\
  result & result & The cluster result \\
  \hline
\end{tabular}
   \end{center}
\end{table}

We will present detailed examples in numerical analysis. For other codes, please see our uploaded R file -- ORKM.R.

\hypertarget{numerical-analysis}{%
\section{Numerical analysis}\label{numerical-analysis}}

\hypertarget{preparation}{%
\subsection{Preparation}\label{preparation}}

The functions \code{ORKMeans} and \code{RKMeans} are used to perform online RKMC and off-line RKMC, respectively. We use three indicators to examine the clustering effect in simulation and in real data analysis. The first indicator is normalized mutual information (NMI), expressed as \[
{\rm NMI}(\Omega;C)=2\frac{I(\Omega;C)}{H(\Omega)+H(C)},
\] where \(I(\Omega;C)\) is the mutual information, \(H(\Omega)+H(C)\) is the normalization information.

The second indicator is Purity. The value of Purity is also between {[}0,1{]}. Purity is defined as \[
\operatorname{Purity}(\Omega,C)=\frac{1}{N}\sum_{k=1}max|w_{k}\bigcap c_{j}|,
\] where \(N\) is the sample size, \(\Omega=\{w_{1},\ldots,w_{K}\}\) represents the classification result, and \(C=\{c_{1},\ldots,c_{J}\}\) represents the division of true class.

The third indicator is F-score, which takes into account the harmonization values of Precision and Recall, and is defined as \[\operatorname{F-score}=\frac{2Precision\times Recall}{Precision+Recall}.\]

For the NMI, \code{method=0} in \code{ORKMeans} and \code{RKMeans} is the command that calculates the NMI values. We also write the function \code{INDEX} to calculate the values of Purity and F-score.

\hypertarget{simulation}{%
\subsection{Simulation}\label{simulation}}

\textcolor{blue}{ For the simulated data from a single perspective, we have developed a data generation environment within the 
$(N,K,\eta,V,t)$ framework. We employed the R package \CRANpkg{ORKM} and other comparable R packages to analyze the data generated in this environment. The results indicate that our R package exhibits superior performance relative to other R packages. It's essential to note, however, that we evaluated only one specific scenario. In different contexts, such as adjusting the value of N, varying the sample size, clusters' size, number of clusters, changing the value of t, and online initial value size, our \CRANpkg{ORKM} package demonstrates robust clustering outcomes. This aspect highlights the motivation behind presenting our simulation data, not to establish the superiority of our package solely based on the data provided in this article. Readers are free to vary the simulation data's set values, and the \CRANpkg{ORKM} package can yield excellent clustering results, affirming its reliability.}

We generate multi/single-view data sets where the data points follow a normal distribution, to examine the performance of the package \CRANpkg{ORKM}.

\textbf{CASE 1. Single-view data}

\begin{figure}
\includegraphics[width=0.5\linewidth]{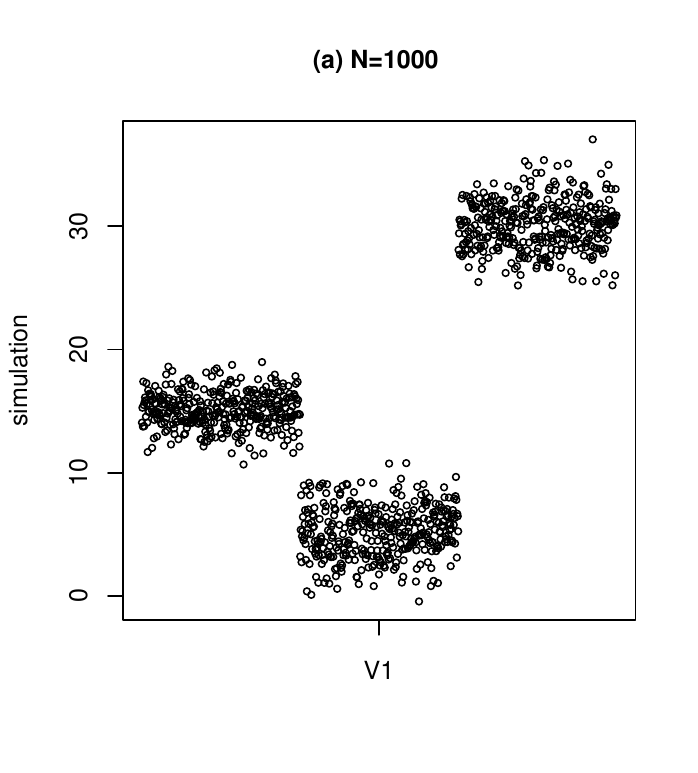} \includegraphics[width=0.5\linewidth]{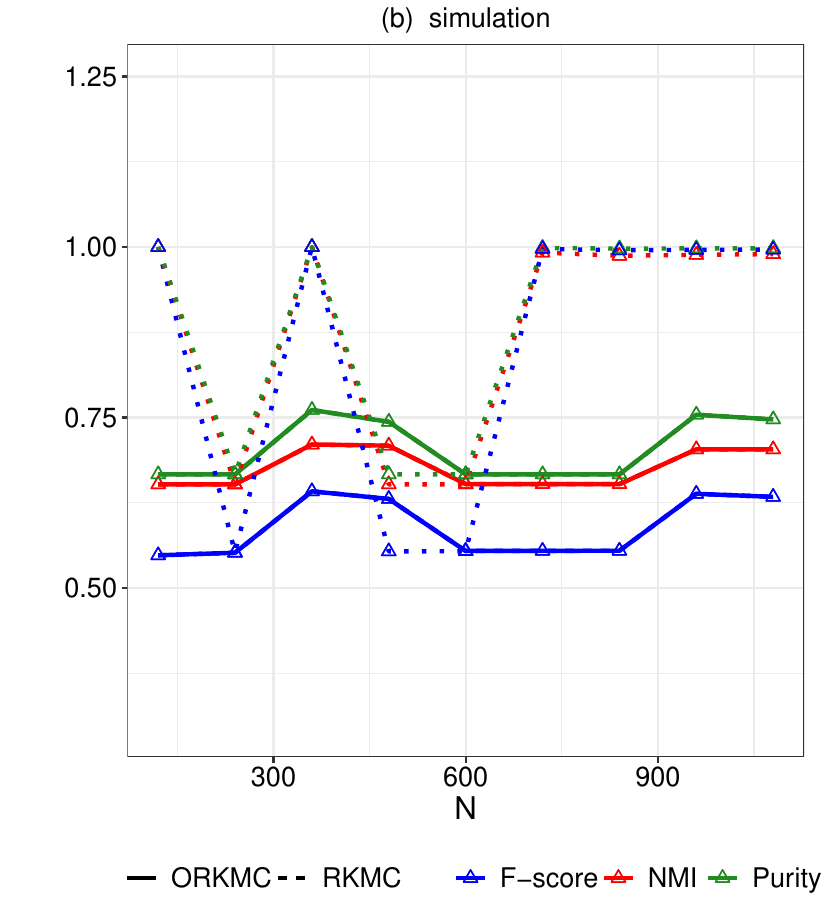} \caption{(a) Simulation data set with $N=1000$;  (b) The stability analysis in single-view data sets}\label{fig:sim}
\end{figure}

Panel (a) in Figure \ref{fig:sim} shows the scatter plot of the single-view simulation data with \((N,K)=(1000,3)\).
\textcolor{blue}{Panel (b) was formed with a fixed set of parameters $(K,J,I)=(3,1,N/2)$ for the ORKMC method, where $N$ lies between 100 and 1000. The performance of the ORKMC method was evaluated based on ${\rm Purity}$ values ranging from 0.6666667 to 0.7611111, $\operatorname{ F-score}$ values ranging from 0.5480226 to 0.6418656, and ${\rm NMI}$ values ranging from 0.6519815 to 0.7104872. Similarly, the RKMC method was evaluated using ${\rm Purity}$ values ranging from 0.6666667 to 1, $\operatorname{ F-score}$ values ranging from 0.5519462 to 1, and ${\rm NMI}$ values ranging from 0.6521382 to 1. The clustering effect of the ORKMC was found to be relatively good when $N$ was equal to 360 or 960, while the clustering effect of the RKMC was consistent with that of the ORKMC when $N$ was equal to 240 or 600. Overall, these results indicate that the package exhibits good stability in single-view simulations.}

\begin{figure}
\includegraphics[width=1\linewidth]{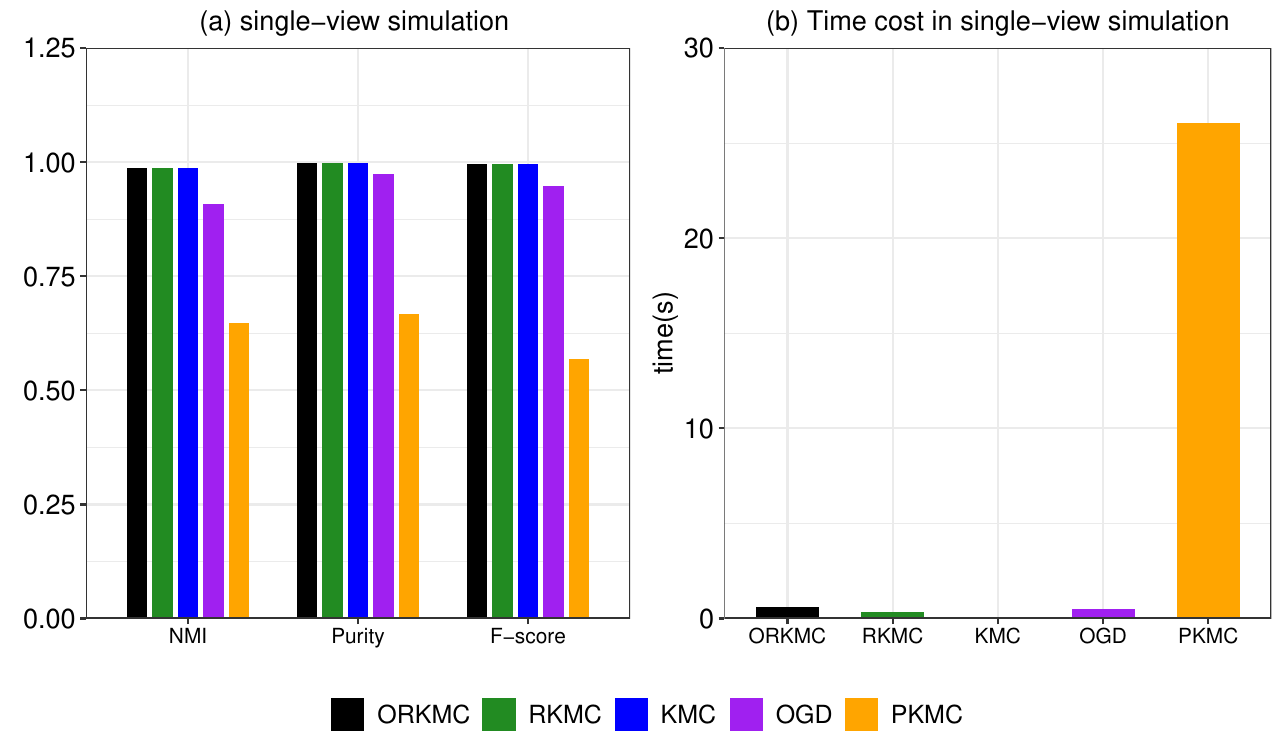} \caption{The comparison results  of the ORKMC in single-view simulation}\label{fig:per1}
\end{figure}

The comparison results in single-view simulation are shown in Figure \ref{fig:per1}.

The results obtained from the experiments are presented in Figure \ref{fig:per1}, Panel (a). \textbackslash textcolor\{blue\}\{For fixed values of \((N,I,K,V,\eta)=(840, 620, 3, 1, 5)\), the ORKMC algorithm exhibited an NMI, Purity, and F-score of 0.987123, 0.997619, and 0.9952212, respectively. Similarly, the RKMC algorithm produced identical scores of 0.987123, 0.997619, and 0.9952212. From Panel (b), we can observe that the time required for running ORKMC and RKMC algorithms were 0.59 and 0.33, respectively. The results indicate that ORKMC and RKMC achieved clustering accuracies close to 100\%. Importantly, the execution time of ORKMC was found to be less than 1 second, suggesting its suitability for large-scale applications.\}

\textbf{CASE 2. Multi-view data}

\begin{figure}
\includegraphics[width=1\linewidth]{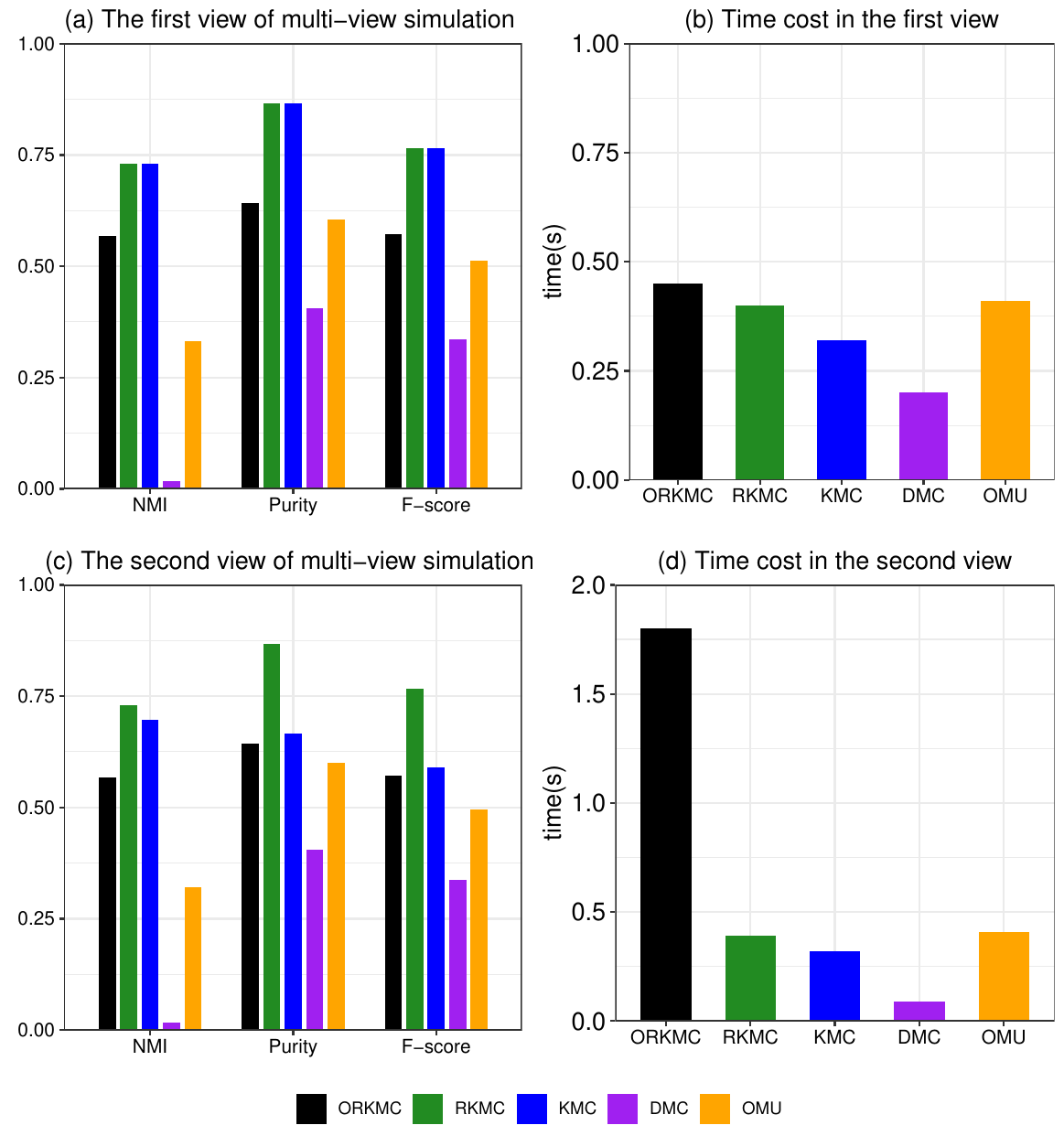} \caption{The performance comparison results  of the ORKMC in multi-view simulation data}\label{fig:perofmulti}
\end{figure}

\begin{figure}
\includegraphics[width=1\linewidth]{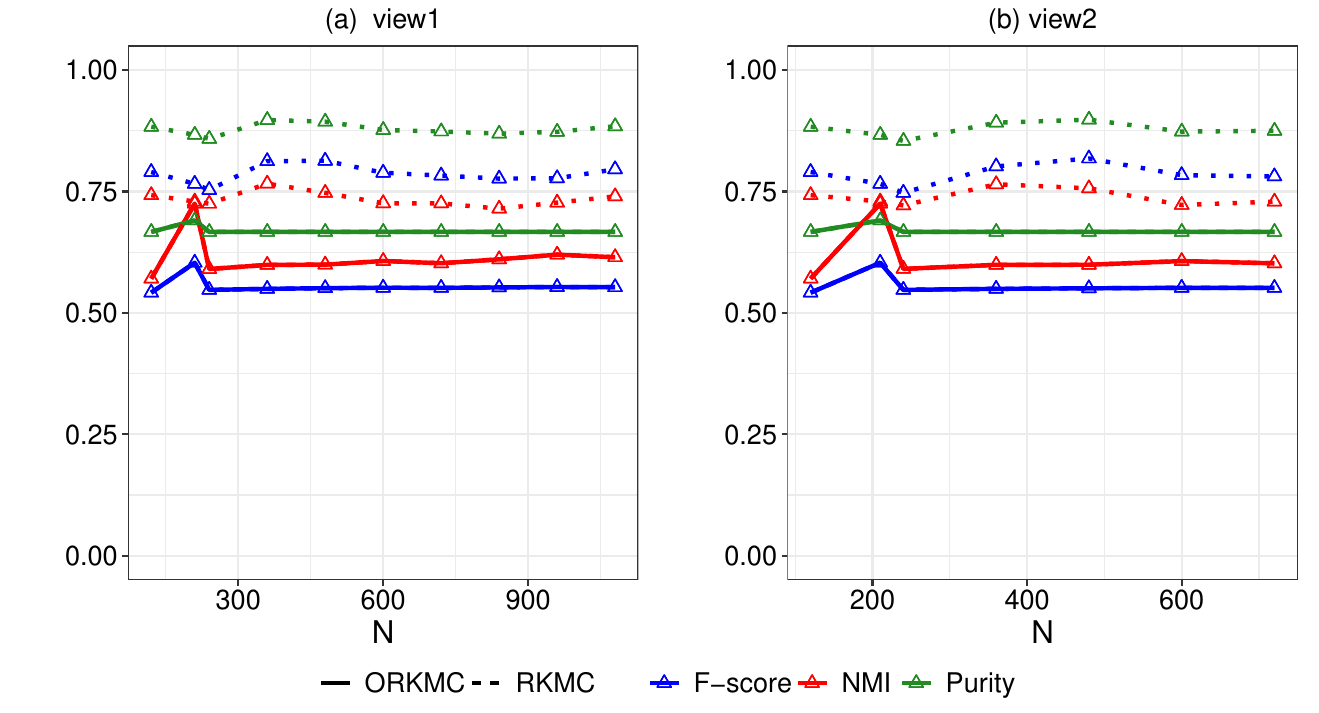} \caption{The stability analysis of the ORKMC}\label{fig:staofmulti}
\end{figure}

We examine the performance of the ORKMC and the RKMC in multi-view simulation, and give a time cost chart of them, see also Figures \ref{fig:perofmulti} and \ref{fig:staofmulti}.

Figure \ref{fig:perofmulti} illustrates the results of multi-view simulations for fixed parameters \((N,I,K,V,\eta)=(210,130,3,2,20)\) in Panels (a) and (c). The ORKMC values obtained from the simulations are 0.56802674, 0.64285714, and 0.57138842 for the first, second, and third views, respectively. In comparison, the RKMC produces values of 0.72928656, 0.86666667, and 0.76581705 for the same views. The time costs of the ORKMC and RKMC are presented in Panels (b) and (d), with the former taking 1.06 and 2.86 units of time for the first and second views, respectively, while the latter taking 0.4 and 0.39 units of time for the same views. The clustering accuracy of the RKMC is found to be higher than that of the DMC by 86\% and 100\%, respectively, whereas the ORKMC as an online method shows a 60\% improvement over the OMU.

\textcolor{blue}{It is important to note that in the real world, there exist various datasets with varying magnitudes. Consequently, for any clustering method, its stability assumes significant importance. In this study, we assess the clustering effect of ORKMC and RKMC concerning different sizes. Figure \ref{fig:staofmulti} presents the findings obtained by fixing $(K,V,J,\eta,I)=(3,2,1,20,N/2)$ while ranging $N \in [100,1000]$, whereby the ORKMC values range from ${\rm Purity} \in [0.6666667,0.6904762]$, $\operatorname{F-score} \in [0.5416164,0.6033130]$, and ${\rm NMI} \in [0.5705238,0.7251129]$ whereas the RKMC values range from  ${\rm Purity} \in [0.8583333,0.8972222]$, $\operatorname{F-score} \in [0.7531177,0.8133381]$, and ${\rm NMI} \in [0.7146113,0.7661900]$. Similarly, for  $N\in[100,720]$ in the second test view, the value ranges remain identical to those in the first view. The ORKM has the most excellent effect when $N=210$. However, for $N>240$, the values do not fluctuate significantly, indicating a stable effect of the ORKMC. Overall, our multi-view simulation results indicate that both the ORKMC and RKMC exhibit good stability levels.}

To examine the performance of the package \CRANpkg{ORKM}, we examine the performance of \CRANpkg{mclust}, \CRANpkg{Spectrum}, and \CRANpkg{ClusterR} for fixed\((N,I,K,\eta)=(210,130,3,5)\). The comparison results are shown in Table \ref{table3}.

\begin{table}[htbp] \small
 \begin{center}
 
  \caption{\textcolor{blue}{The comparison  results of the package \CRANpkg{ORKM} in simulation}} \label{table3}

  \begin{tabular}{ c c c c c c}  
  \hline
  data set & packages & NMI & Purity & F-score& time(s) \\ 
 \hline
  single-view & mclust & \textbf{0.9928104} & \textbf{0.9988095}  & \textbf{0.9976105}& 0.72\\
 & Spectrum & 0.6520966 & 0.6666667 & 0.5545859& 0.86\\
 & ClusterR & 0.6520966 & 0.6666667 & 0.5545859& 1.34 \\
 & stats & 0.987123 & 0.9976190 & 0.9952212 & 0.9 \\
 & fpcCBI & 0.987123 & 0.9976190 & 0.9952212 & 0.76 \\
 & fpcruns & 0.987123 & 0.9976190 & 0.9952212 & 1.1 \\
 & ORKM & 0.9871230 &  0.9976190 & 0.9952212& 0.33\\
  \hline
 multi-view 1 & mclust & 0.7227361 & 0.8333333 & 0.7141829& 0.19\\
  & Spectrum & 0.7249822 & 0.8619048 & 0.7589401& 0.24  \\
  & ClusterR & 0.7208473 & 0.8571429 & 0.7522056& 0.09 \\
   & stats & 0.7249822 & 0.8619048 & 0.7589401 & 0.89 \\
 & fpcCBI & 0.7249822 & 0.8619048 & 0.7589401 & 0.66 \\
 & fpcruns & 0.7249822 & 0.8619048 & 0.7589401 & 1.5 \\
  & ORKM & \textbf{0.7292866} & \textbf{0.8666667} & \textbf{0.7658171}& 0.36\\
     \hline
multi-view 2  & mclust & 0.7227361 & 0.8333333 & 0.7141829&  0.14 \\
  & Spectrum & 0.7249822 & 0.8619048 & 0.7589401& 0.23\\
  & ClusterR & 0.7208473 & 0.8571429 &  0.7522056&  0.15\\
     & stats & 0.7249822 & 0.8619048 & 0.7589401 & 0.89 \\
 & fpcCBI & 0.7249822 & 0.8619048 & 0.7589401 & 0.66 \\
 & fpcruns & 0.7249822 & 0.8619048 & 0.7589401 & 1.5 \\
  & ORKM & \textbf{0.7292866} & \textbf{0.8666667} &  \textbf{ 0.7658171}& 1.69\\
    \hline
\end{tabular}
   \end{center}
\end{table}

\textcolor{blue}{Based on the results displayed in Table \ref{table3}, it is evident that \CRANpkg{mclust} exhibits the best performance in single-view simulation. However, in multi-view simulation scenarios, \CRANpkg{ORKM} consistently demonstrates superior clustering accuracy compared to the other three packages, regardless of whether it is applied to the first view or the second view. Therefore, \CRANpkg{ORKM} can be regarded as the package with the highest clustering accuracy among the tested ones. The second-highest performing package in the multi-view simulation is \CRANpkg{Spectrum}.}

\hypertarget{real-data-analysis}{%
\subsection{Real data analysis}\label{real-data-analysis}}

In the section, we examine the performance of the package \CRANpkg{ORKM} in real data sets.

\hypertarget{qcm-data-set}{%
\subsubsection{QCM data set}\label{qcm-data-set}}

QCM \citep{QCM} data set is a single-view data set with \(K=5\). The data set is publicly available from the UCI repository, and available with the package by using the following command, R code in QCM data set is expressed as

\begin{verbatim}
data(QCM)
QCM1 <- QCM[,-11:-15]; QCM1[1:25,11] <- 1; QCM1[26:50,11] <- 2
QCM1[51:75,11] <- 3; QCM1[76:100,11] <- 4; QCM1[101:125,11] <- 5
data <- QCM1[,-11]; X <- as.matrix(data)
N <- nrow(X); J <- ncol(X); N; J
\end{verbatim}

\begin{verbatim}
#> [1] 125
\end{verbatim}

\begin{verbatim}
#> [1] 10
\end{verbatim}

\begin{verbatim}
set.seed(123)
ptm <- proc.time()
q2 = RKMeans(X = X, K = 5, V = 1, yita = 110, r = 0.5, max.iter = 1000, truere = QCM1[,11], 
             method = 0)
timeRKM <- proc.time()-ptm
purityRKM <- INDEX(q2$result, QCM1[,11], method = 0)
FRKM <- INDEX(q2$result, QCM1[,11], method = 3)
list(timeRKM, q2$NMI, purityRKM, FRKM)
\end{verbatim}

\begin{verbatim}
#> [[1]]
#>    user  system elapsed 
#>    0.13    0.02    0.22 
#> 
#> [[2]]
#> [1] 0.2912228
#> 
#> [[3]]
#> [1] 0.464
#> 
#> [[4]]
#> [1] 0.3194955
\end{verbatim}

The result shows that QCM data set includes 125 instances on 11 variables, and the 11-th column is the real label. For easy calculation, we can replace the labels with the numbers 1, 2, 3, 4, and 5, while we also need to remove the labels before clustering.

\begin{figure}
\includegraphics[width=1\linewidth]{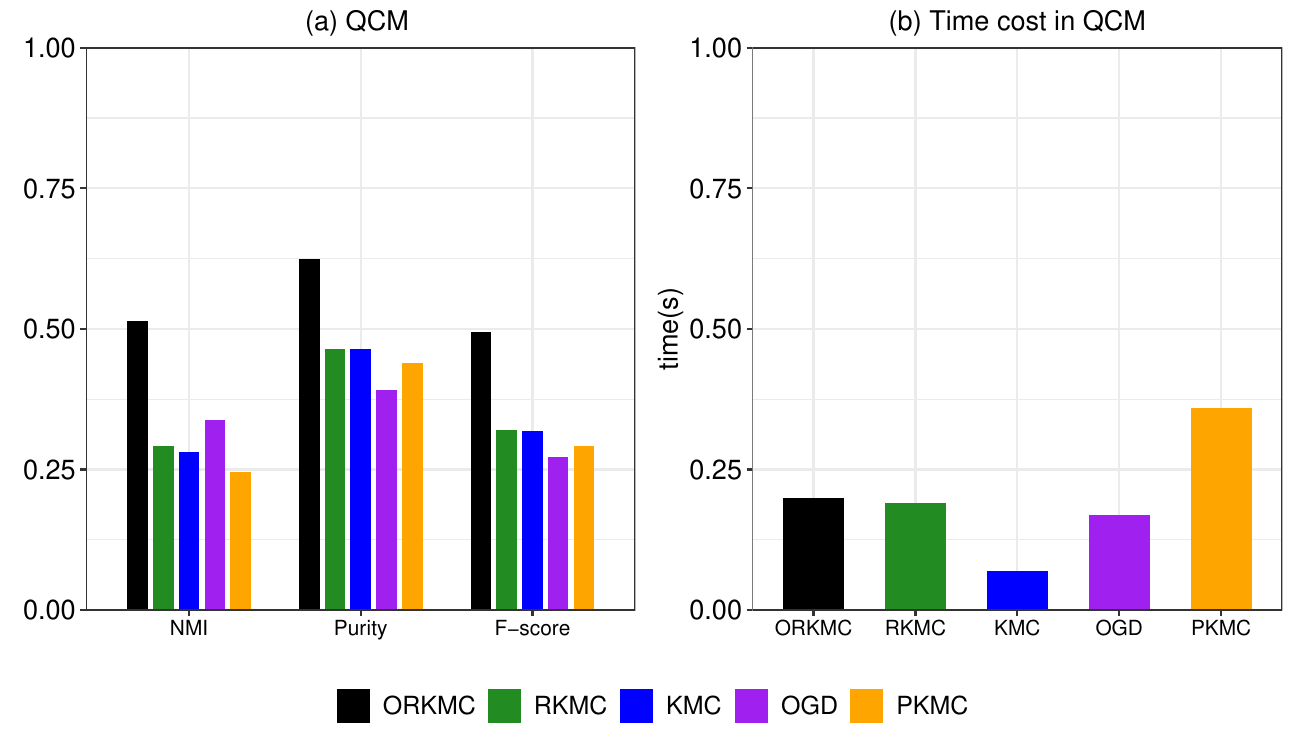} \caption{The   comparison results  of the ORKMC in QCM data set}\label{fig:perofqcm}
\end{figure}

With \CRANpkg{ggplot2} \citep{ggplot2}, we present the comparison results of the ORKMC in Figure \ref{fig:perofqcm}. The following paragraph provides an analysis of the performance metrics and computational efficiency for two clustering algorithms, ORKMC and RKMC, on the QCM dataset. Referring to Panel (a) in Figure \ref{fig:perofqcm}, the NMI, Purity, and F-score values achieved by ORKMC were 0.5137041, 0.624, and 0.4939271, respectively. In contrast, the corresponding values for RKMC were lower, at 0.2912228, 0.464, and 0.3194955, respectively. Moving on to Panel (b), it can be observed that the time costs of ORKMC and RKMC were 0.09 and 0.18, respectively. Notably, the clustering effectiveness of ORKMC was relatively higher compared to that of RKMC, which ranks second. This is evidenced by an improvement in NMI values by 76\%, Purity values by 34\%, and F-score values by 58\%.

\hypertarget{movie-data-set}{%
\subsubsection{Movie data set}\label{movie-data-set}}

Movie data set as a multi-view data set includes 2 views, each view includes 617 instances of 1878 variables with \(K=17\). The data set is available in the package \CRANpkg{ORKM}, and expressed as

\begin{verbatim}
data(movie_1); data(movie_2); data(turelabel); label0 <- as.vector(turelabel)
label0<-unlist(turelabel); label<-c(label0)
X1 <- as.matrix(movie_1); X2 <- as.matrix(movie_2)
N1 <- nrow(X1); J1 <- ncol(X1); N2 <- nrow(X2); J2 <- ncol(X2)
set.seed(123)
ptm <- proc.time()
vq1 = ORKMeans(X = X1, K = 17, V = 2, chushi = 600, r = 0.5, gamma = 1e-05,
               epsilon = 1, yita = 0.5, truere = label, max.iter = 10 , method = 0)
v1timeORKM <- proc.time()-ptm
v1purityORKM <- INDEX(vq1$result, label, method = 0)
v1FORKM <- INDEX(vq1$result, label, method = 3)
list(v1timeORKM, vq1$NMI, v1purityORKM, v1FORKM)
\end{verbatim}

\begin{verbatim}
#> [[1]]
#>    user  system elapsed 
#>   11.07    0.30   26.10 
#> 
#> [[2]]
#> [1] 0.2158074
#> 
#> [[3]]
#> [1] 0.2106969
#> 
#> [[4]]
#> [1] 0.09525322
\end{verbatim}

\begin{figure}
\includegraphics[width=1\linewidth]{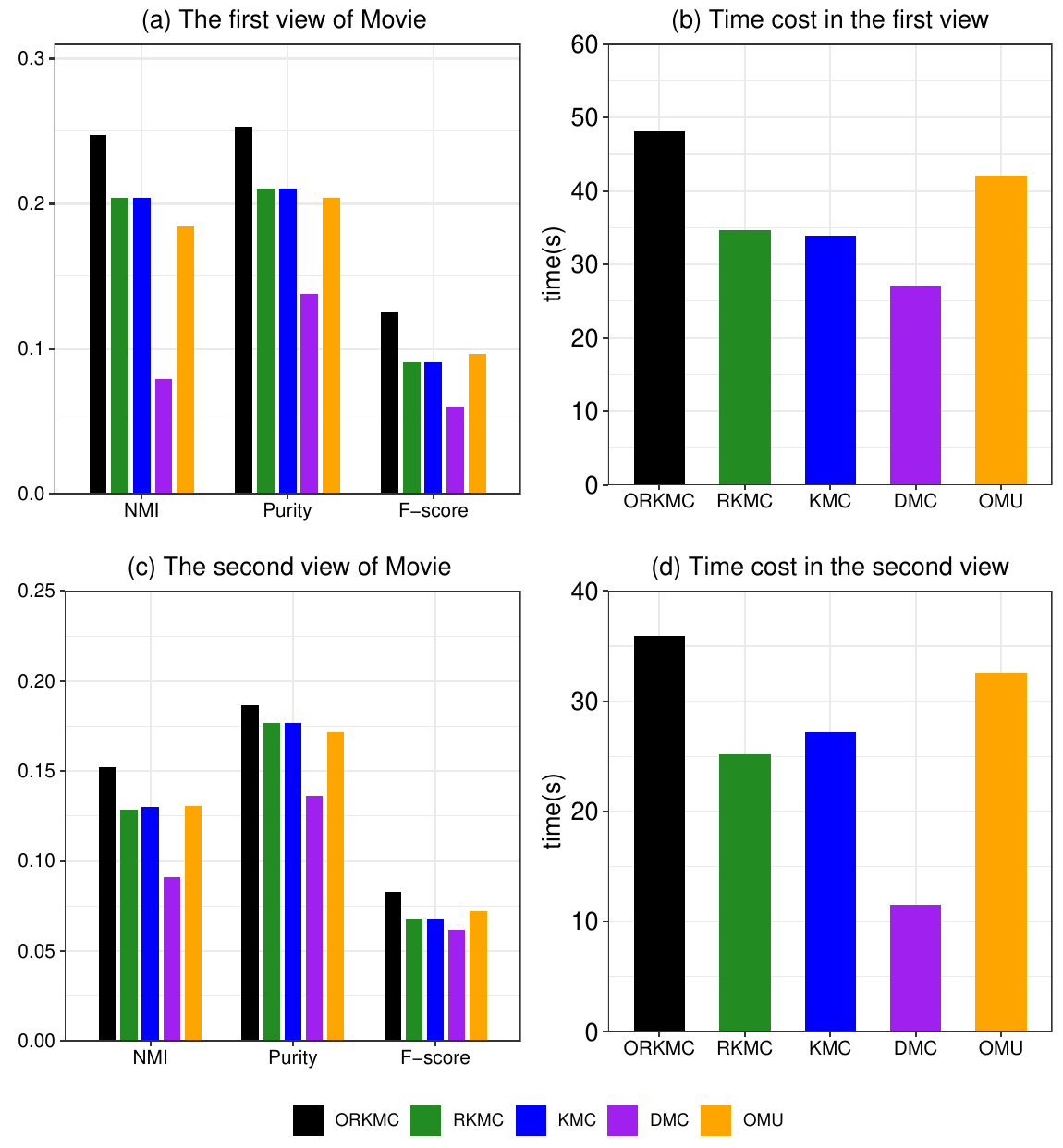} \caption{The comparison results of the ORKMC  in Movie data set}\label{fig:perofmov}
\end{figure}

With \CRANpkg{ggplot2} \citep{ggplot2} in Movie data set, we present the comparison results of the ORKMC in Figure \ref{fig:perofmov}.

Panel (a) in Figure \ref{fig:perofmov} demonstrates the ORKMC values in view 1 of the Movie dataset, which are 0.2476449, 0.2528363, and 0.12482790, respectively. In comparison, the RKMC values for this dataset are 0.2042879, 0.2106969, and 0.09102353, respectively. The time costs of the ORKMC and RKMC, shown in Panel (b), are 48.42 and 33.15, respectively. Specifically, the ORKMC showed a higher clustering effect than the RKMC, as indicated by a 20\% increase in NMI values, 19\% increase in Purity values, and 33\% increase in F-score values. Similarly, in Panel (c), the ORKMC and RKMC values for View 2 are 0.1522948, 0.1863857, 0.08292683 and 0.1285667, 0.1766613, 0.06796304, respectively. Moreover, the time costs of ORKMC and RKMC in Panel (d) are 26.48 and 24.63, respectively. Again, the ORKMC method performed better than the RKMC method in terms of clustering effectiveness, with NMI values increasing by 25\%, Purity values increasing by 5\%, and F-score values increasing by 33\%.

The comparison results of the ORKMC with different packages in real data set are shown in Table \ref{table4}.

\begin{table}[htbp] \small
  \centering
    \caption{\textcolor{blue}{The comparison results of the package \CRANpkg{ORKM} in real data sets}}\label{table4}
  \begin{tabular}{ c c c c c c }
  \hline
  data set & packages & NMI & Purity & F-score & time(s)  \\
        \hline
 QCM & mclust & 0.3728653 & 0.44 & 0.36363 &  0.26\\
  & Spectrum & 0.3134887 & 0.472 & 0.3522655 &0.16 \\
  & ClusterR & 0.262914 &0.488 & 0.3077726  &0.09 \\
  & stats & 0.2912228 & 0.464& 0.3194955&0.16 \\
  &fpcCBI & 0.2912228 &0.464 & 0.3194955& 0.2 \\
  &fpcruns & 0.215539 & 0.424 & 0.2920679 & 0.3 \\
  & ORKM & \textbf{0.5137041} & \textbf{0.624} & \textbf{0.4939271} & 0.07 \\
  \hline
Movie view 1  & mclust & 0.1629454 & 0.1426256 & 0.06419356 & 4.53\\
  & Spectrum & \textbf{0.2581707} & 0.2398703 & 0.08346585 &0.95 \\
  & ClusterR & 0.1514069 & 0.1523501 & 0.06110052  & 4.37\\
  & stats & 0.2276991 & 0.2398703& 0.1125513&6.16 \\
  &fpcCBI & 0.2276991 &0.2398703 & 0.1125513& 8.35 \\
  &fpcruns & 0.232048 & 0.2366288 &0.08765364 &7.96 \\
  & ORKM & 0.2476449 & \textbf{0.2528363} & \textbf{0.12482790} &2.15 \\
    \hline
Movie view 2  & mclust & 0.1601179 & 0.1426256 & 0.06527764 &1.89 \\
  & Spectrum & \textbf{0.1811338} & 0.165316 & 0.06670727 & 0.65 \\
  & ClusterR & 0.108323 & 0.1037277 & 0.06109474& 4.44   \\
  & stats & 0.1652317 & 0.1620746& 0.06963312&5.16 \\
  &fpcCBI & 0.1652317 &0.1620746 & 0.06963312& 7.85 \\
  &fpcruns & 0.1774326 & 0.1572123 &0.06582003 &7.46 \\
  & ORKM & 0.1522948 & \textbf{0.1863857} &  \textbf{0.08292683}& 3.18 \\
    \hline
    \end{tabular}
\end{table}

As shown in Table \ref{table4}, the \CRANpkg{ORKM} package demonstrates superior performance on the QCM dataset. Notably, its clustering effect on Purity exceeds that of other packages, with a 41\% increase compared to \CRANpkg{mclust}, a 31\% increase compared to \CRANpkg{Spectrum}, and a 27\% increase compared to \CRANpkg{ClusterR}. Similarly, when evaluating the multi-view Movie dataset, the \CRANpkg{ORKM} package remains the top-performing solution. For the first view, it outperforms \CRANpkg{Spectrum} by 4\%, \CRANpkg{mclust} by 31\%, and \CRANpkg{ClusterR} by 66\%. On the second view, the clustering effect of \CRANpkg{ORKM} exceeds other packages with a 28\% advantage compared to \CRANpkg{mclust}, a 12.5\% advantage compared to \CRANpkg{Spectrum}, and an 80\% advantage compared to \CRANpkg{ClusterR}.

\hypertarget{summary}{%
\section{Summary}\label{summary}}

\textcolor{blue}{The reason why ORKMC is called a multi-view data solver is that, at the beginning of ORKMC, we used a model suitable for multi-view data as a starting point, and a method suitable for updating multi-view data, while ensuring clustering accuracy and speed. Due to the specificity of multi-view data, each view can only represent part of the true characteristics of the data. Based on this principle, views can be divided into master views and non-master views, with master views representing more than 60\% of the characteristics of the data. However, the process of finding the master view is unknown and ORKMC updates the weights of each view by assigning initial values to each view and by calculating the clustering indicator matrix and the clustering centre matrix each time. This is done until the clustering is complete, i.e. the range of the two changes before and after the cluster centre matrix is less than a threshold value. From there ORKMC obtains the weight values for the unknown dataset by calculation. The data recorded by a multi-view view all serve the same dataset. However, we can use ORKMC to calculate the weights for each view of the multi-view data to evaluate the data integrity of the views and thus make changes accordingly when clustering. In summary, ORKMC is called a multi-view problem solver because, on the one hand, the ORKMC algorithm can calculate the clustering results for each view of the data and has better robustness and effectiveness relative to other multi-view algorithms. On the other hand, the ORKMC algorithm is widely used because it can handle multi-view data, representing a larger number of meaningful data sets, than other clustering algorithms. Either way, ORKMC can both handle multi-view datasets and has good clustering accuracy on multi-view datasets. This is why it is justifiably called a multi-view problem solver.}

\textcolor{blue}{The proposed \CRANpkg{ORKM} package exhibits higher clustering accuracy than other commonly used function packages. In numerical analysis, the efficacy of ORKMC's clustering potential can be influenced by several factors such as sample size, number of categories, regularization parameter, and balance parameter. Through a multi/single-view simulation, we evaluated the clustering ability of ORKMC and found that it exhibited better stability for sample sizes within a particular range. Notably, on the QCM dataset, ORKMC displayed the best performance, followed by RKMC, KMC, OGD, and PKMC in that order. Similarly, on the multi-view Movie dataset, ORKMC and RKMC were the top two performers, whereas OMU, KMC, and DMC occupied the third to fifth positions in respective order. Additionally, when benchmarked against other packages, \CRANpkg{ORKM} presented superior clustering effects under the studied scenarios.}

\textcolor{blue}{It is worth noting that the ORKM incurs the highest computational time when only one sample is updated at each time node. Alternatively, updating one data block at a time represents an easier approach. Furthermore, several distinguished researchers have developed numerous packages for model selection. Combining online multi-view update and model selection represents a very promising avenue for future research.}

\hypertarget{references}{%
\section{References}\label{references}}

\bibliography{ORKM.bib}

\begin{thebibliography}{34}
\providecommand{\natexlab}[1]{#1}
\providecommand{\url}[1]{\texttt{#1}}
\expandafter\ifx\csname urlstyle\endcsname\relax
  \providecommand{\doi}[1]{doi: #1}\else
  \providecommand{\doi}{doi: \begingroup \urlstyle{rm}\Url}\fi

\bibitem[Adak et~al.(2020)Adak, Lieberzeit, Jarujamrus, and Yumusak]{QCM}
M.~F. Adak, P.~Lieberzeit, P.~Jarujamrus, and N.~Yumusak.
\newblock Classification of alcohols obtained by qcm sensors with different
  characteristics using abc based neural network.
\newblock \emph{Engineering Science and Technology, an International Journal},
  23\penalty0 (3):\penalty0 463--469, 2020.

\bibitem[Cai et~al.(2013)Cai, Nie, and Huang]{MKM}
X.~Cai, F.~Nie, and H.~Huang.
\newblock Multi-view k-means clustering on big data.
\newblock In \emph{Twenty-Third International Joint conference on artificial
  intelligence}, 2013.

\bibitem[Chao et~al.(2018)Chao, Sun, and Bi]{chao2018}
G.~Chao, S.~Sun, and J.~Bi.
\newblock A survey on multi-view clustering.
\newblock \emph{IEEE Transactions on Artificial Intelligence}, 2\penalty0
  (2):\penalty0 146--168, 2018.

\bibitem[Ding et~al.(2005)Ding, He, and Simon]{Ding2005}
C.~Ding, X.~He, and H.~D. Simon.
\newblock Nonnegative lagrangian relaxation of k-means and spectral clustering.
\newblock In \emph{European Conference on Machine Learning}, pages 530--538.
  Springer, 2005.

\bibitem[Feng et~al.(2013)Feng, Xu, and Yan]{feng2013online}
J.~Feng, H.~Xu, and S.~Yan.
\newblock Online robust pca via stochastic optimization.
\newblock \emph{Advances in Neural Information Processing Systems},
  26:\penalty0 404–412, 2013.

\bibitem[Guo et~al.(2022)Guo, Yu, Song, and Niu]{ORKM}
G.~Guo, M.~Yu, H.~Song, and R.~Niu.
\newblock \emph{ORKM: The Online Regularized K-Means Clustering Algorithm},
  2022.
\newblock URL \url{https://CRAN.R-project.org/package=ORKM}.
\newblock R package version 0.0.2.1.

\bibitem[Guo et~al.(2014)Guo, Hao, and Liu]{guo2014incremental}
L.~Guo, J.~H. Hao, and M.~Liu.
\newblock An incremental extreme learning machine for online sequential
  learning problems.
\newblock \emph{Neurocomputing}, 128:\penalty0 50--58, 2014.

\bibitem[Hennig(2023)]{fpc}
C.~Hennig.
\newblock \emph{fpc: Flexible Procedures for Clustering}, 2023.
\newblock URL \url{https://CRAN.R-project.org/package=fpc}.
\newblock R package version 2.2-10.

\bibitem[Hoi et~al.(2021)Hoi, Sahoo, Lu, and Zhao]{OGD}
S.~C. Hoi, D.~Sahoo, J.~Lu, and P.~Zhao.
\newblock Online learning: A comprehensive survey.
\newblock \emph{Neurocomputing}, 459:\penalty0 249--289, 2021.

\bibitem[John and Watson(2020)]{Spectrum}
C.~R. John and D.~Watson.
\newblock \emph{Spectrum: Fast Adaptive Spectral Clustering for Single and
  Multi-View Data}, 2020.
\newblock URL \url{https://CRAN.R-project.org/package=Spectrum}.
\newblock R package version 1.1.

\bibitem[Kim et~al.(2018)Kim, Han, and Whang]{kim2018}
J.~Kim, Y.~Han, and I.~H. Whang.
\newblock Online $l\_1$ regularized optimization for optical misalignment
  cancellation of strapdown seekers.
\newblock \emph{International Journal of Aeronautical and Space Sciences},
  19\penalty0 (3):\penalty0 711--717, 2018.

\bibitem[Kumar et~al.(2011)Kumar, Rai, and Daum\'{e}]{kumar2011co}
A.~Kumar, P.~Rai, and H.~Daum\'{e}.
\newblock Co-regularized multi-view spectral clustering.
\newblock In \emph{Proceedings of the 24th International Conference on Neural
  Information Processing Systems}, page 1413–1421. Curran Associates Inc.,
  2011.

\bibitem[Lee and Seung(1999)]{lee1999learning}
D.~D. Lee and H.~S. Seung.
\newblock Learning the parts of objects by non-negative matrix factorization.
\newblock \emph{Nature}, 401\penalty0 (6755):\penalty0 788--791, 1999.

\bibitem[Lee et~al.(2020)Lee, Wang, and Schifano]{lee2020}
J.~Lee, H.~Wang, and E.~D. Schifano.
\newblock Online updating method to correct for measurement error in big data
  streams.
\newblock \emph{Computational Statistics and Data Analysis}, 149:\penalty0
  106976, 2020.

\bibitem[Liberty et~al.(2016)Liberty, Sriharsha, and Sviridenko]{OKM}
E.~Liberty, R.~Sriharsha, and M.~Sviridenko.
\newblock An algorithm for online k-means clustering.
\newblock In \emph{2016 Proceedings of the eighteenth workshop on algorithm
  engineering and experiments}, pages 81--89. SIAM, 2016.

\bibitem[Lu et~al.(2016)Lu, Yan, and Lin]{lu2016}
C.~Lu, S.~Yan, and Z.~Lin.
\newblock Convex sparse spectral clustering: Single-view to multi-view.
\newblock \emph{IEEE Transactions on Image Processing}, 25\penalty0
  (6):\penalty0 2833--2843, 2016.

\bibitem[MacQueen(1967)]{MacQueen1967}
J.~MacQueen.
\newblock Some methods for classification and analysis of multivariate
  observations.
\newblock 1967.

\bibitem[Mouselimis(2022)]{ClusterR}
L.~Mouselimis.
\newblock \emph{{ClusterR}: Gaussian Mixture Models, K-Means,
  Mini-Batch-Kmeans, K-Medoids and Affinity Propagation Clustering}, 2022.
\newblock URL \url{https://CRAN.R-project.org/package=ClusterR}.
\newblock R package version 1.2.6.

\bibitem[Scrucca et~al.(2016)Scrucca, Fop, Murphy, and Raftery]{mclust}
L.~Scrucca, M.~Fop, T.~B. Murphy, and A.~E. Raftery.
\newblock {mclust} 5: clustering, classification and density estimation using
  {G}aussian finite mixture models.
\newblock \emph{The {R} Journal}, 8\penalty0 (1):\penalty0 289--317, 2016.
\newblock URL \url{https://doi.org/10.32614/RJ-2016-021}.

\bibitem[Seung and Lee(2001)]{OMU}
D.~Seung and L.~Lee.
\newblock Algorithms for non-negative matrix factorization.
\newblock \emph{Advances in neural information processing systems},
  13:\penalty0 556--562, 2001.

\bibitem[Shao et~al.(2016)Shao, He, Lu, and Philip]{shao2016b}
W.~Shao, L.~He, C.~T. Lu, and S.~Y. Philip.
\newblock Online multi-view clustering with incomplete views.
\newblock In \emph{2016 IEEE International conference on big data}, pages
  1012--1017. IEEE, 2016.

\bibitem[Wang et~al.(2007)Wang, Chen, and Li]{wang2007group}
L.~Wang, G.~Chen, and H.~Li.
\newblock Group scad regression analysis for microarray time course gene
  expression data.
\newblock \emph{Bioinformatics}, 23\penalty0 (12):\penalty0 1486--1494, 2007.

\bibitem[Wei et~al.(2020)Wei, Wang, Yu, Domeniconi, and Zhang]{DMC}
S.~Wei, J.~Wang, G.~Yu, C.~Domeniconi, and X.~Zhang.
\newblock Multi-view multiple clusterings using deep matrix factorization.
\newblock In \emph{Proceedings of the AAAI conference on artificial
  intelligence}, volume~34, pages 6348--6355, 2020.

\bibitem[White et~al.(2012)White, Zhang, Schuurmans, and Yu]{white2012}
M.~White, X.~Zhang, D.~Schuurmans, and Y.~l. Yu.
\newblock Convex multi-view subspace learning.
\newblock In \emph{Advances in Neural Information Processing Systems},
  volume~25. Curran Associates, Inc., 2012.

\bibitem[Wickham(2016)]{ggplot2}
H.~Wickham.
\newblock \emph{ggplot2: Elegant Graphics for Data Analysis}.
\newblock Springer-Verlag New York, 2016.
\newblock URL \url{https://ggplot2.tidyverse.org}.

\bibitem[Xu and Lange(2019)]{PKM}
J.~Xu and K.~Lange.
\newblock Power k-means clustering.
\newblock In \emph{International conference on machine learning}, pages
  6921--6931, 2019.

\bibitem[Yang et~al.(2020)Yang, Zhang, and Tang]{yang2021a}
L.~Yang, L.~Zhang, and Y.~Tang.
\newblock Online binary incomplete multi-view clustering.
\newblock In \emph{Joint European Conference on Machine Learning and Knowledge
  Discovery in Databases}, pages 75--90. Springer, 2020.

\bibitem[Yin et~al.(2013)Yin, Chen, and Hu]{softKM}
X.~Yin, S.~Chen, and E.~Hu.
\newblock Regularized soft k-means for discriminant analysis.
\newblock \emph{Neurocomputing}, 103:\penalty0 29--42, 2013.

\bibitem[Zhang(2010)]{zhang2010nearly}
C.~H. Zhang.
\newblock Nearly unbiased variable selection under minimax concave penalty.
\newblock \emph{The Annals of statistics}, 38\penalty0 (2):\penalty0 894--942,
  2010.

\bibitem[Zhao et~al.(2014)Zhao, Evans, and Dugelay]{zhao2014}
X.~Zhao, N.~Evans, and J.~L. Dugelay.
\newblock A subspace co-training framework for multi-view clustering.
\newblock \emph{Pattern Recognition Letters}, 41:\penalty0 73--82, 2014.

\bibitem[Zhou et~al.(2014)Zhou, Liu, and Zhu]{zhou2014}
X.~Zhou, Z.~Liu, and C.~Zhu.
\newblock Online regularized and kernelized extreme learning machines with
  forgetting mechanism.
\newblock \emph{Mathematical Problems in Engineering}, pages 1--11, 2014.

\bibitem[Zong et~al.(2017)Zong, Zhang, Zhao, Yu, and Zhao]{zong2017}
L.~Zong, X.~Zhang, L.~Zhao, H.~Yu, and Q.~Zhao.
\newblock Multi-view clustering via multi-manifold regularized non-negative
  matrix factorization.
\newblock \emph{Neural Networks}, 88:\penalty0 74--89, 2017.

\bibitem[Zou(2006)]{zou2006adaptive}
H.~Zou.
\newblock The adaptive lasso and its oracle properties.
\newblock \emph{Journal of the American statistical association}, 101\penalty0
  (476):\penalty0 1418--1429, 2006.

\bibitem[Zou and Hastie(2005)]{zou2005regularization}
H.~Zou and T.~Hastie.
\newblock Regularization and variable selection via the elastic net.
\newblock \emph{Journal of the royal statistical society: series B (statistical
  methodology)}, 67\penalty0 (2):\penalty0 301--320, 2005.

\end{thebibliography}

\address{%
Miao Yu\\
Shandong University of Technology\\%
School of Mathematics and Statistics\\ Zibo, China\\
\href{mailto:miaoyusdut@gmail.com}{\nolinkurl{miaoyusdut@gmail.com}}%
}

\address{%
Guangbao Guo\\
Shandong University of Technology\\%
School of Mathematics and Statistics\\ Zibo, China\\
\textit{ORCiD: \href{https://orcid.org/0000-0002-4115-6218}{0000-0002-4115-6218}}\\%
\href{mailto:ggb11111111@sdut.edu.cn}{\nolinkurl{ggb11111111@sdut.edu.cn}}%
}

\end{article}

\end{document}